\documentclass[multphys,vecphys,a4paper,natbib]{svmult}
\usepackage{graphicx}        
\usepackage{multicol}        
\usepackage[bottom]{footmisc}
\usepackage{journals}

\begin{document}

\title*{Asteroseismology across the HR diagram}
\author{M\'ario~J.~P.~F.~G.~Monteiro\inst{1,2}}
\institute{Centro de Astrof\'{\i}sica da Universidade do Porto,
   Rua das Estrelas, 4150-762 Porto, Portugal
(\texttt{mjm@astro.up.pt})
\and Departamento de Matem\'atica Aplicada,
   Faculdade de Ci\^encias da Universidade do Porto, Portugal}
\maketitle

High precision spectroscopy provides essential information necessary to fully exploit the opportunity of probing the internal structure of stars using Asteroseismology.
In this work we discuss how {\em Asteroseismology} combined with {\em High Precision Spectroscopy} can establish a detailed view on stellar structure and evolution of stars across the HR diagramme.

\section{Introduction}

Stars are fundamental units of the visible Universe, making stellar structure and evolution a conner stone of Astrophysics.
However, the modelling of stellar structure and evolution still has quite a few fundamental questions to address and resolve.
Among these we have convection, atomic diffusion and non-standard effects(like rotation and additional mixing processes).

In order to reproduce with models the observables for a star there is
a collection of parameters that must be choosen.
The mathematical solution is then adjusted with the parameters to a limited set of available boundary conditions provided by the observations.
An improvement or test of the modelling is only possible if the number of parameters is smaller than the observables.
This is usually not the case, so complementary observational information is required to effectively constrain the models and the physics used to calculate these.

\begin{figure}
\centering
\includegraphics[width=8cm]{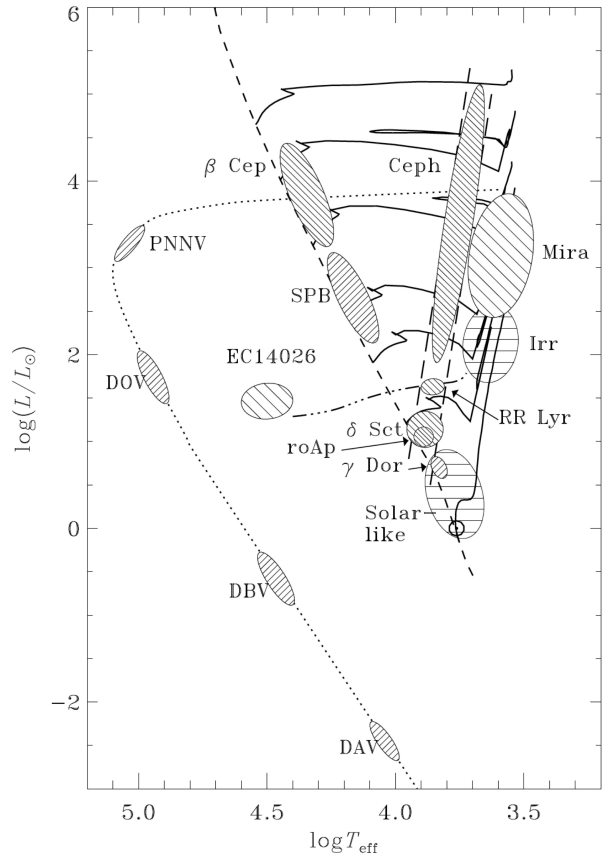}
\caption{Asteroseismic HR Diagram from \citet{cd99} illustrating some of the known pulsators.
The study of solar-like pulsators have benefit greatly from the tools developed for Helioseismology -- seismic study of the Sun -- which are now being applied to analyse ground and space observations of other stars.}
\label{mjm-fig:hrdig}
\end{figure}

To achieve this we must combine all available observational techniques that can provide information on specific stars.
Besides the ``classical'' observables provided by photometry, spectroscopy
and astrometry we now have the possibility to use interferometry (e.g. \citealt{mjm-pijpers03,mjm-difolco04}) and {\em Asteroseismology}.
In particular the detection of periodic variations (oscillations) in most stars -- the field known as Asteroseismology -- has provide us with a powerful tool to analyse the inner structure of stars (for a detailed review see \citealt{mjm-cd04,mjm-cd07} and references therein).
The potential is great as almost all stars across the HR diagram have been found to display oscillations (some of these are identified in Fig.~\ref{mjm-fig:hrdig}).

In this work we will discuss briefly how these stellar oscillations are use to provide additional observational constraints on the models.
We will concentrate on some examples of Asteroseismology with the potential for providing tests that can help to understand some of the fundamental problems in stellar structure and evolution.

Considering the topic of this volume we also give special attention to the benefit of having precise spectroscopic information to produce conclusive seismic tests on specific pulsators in the HR diagram.
Spectroscopy can be used in two ways; either by providing precise frequency determinations (and amplitudes) or by allowing a precise determination of global stellar characteristics.
In either case the quality of the seismic inference is greatly improved by the quality of the spectroscopy measurements done as already shown by some of the currently available facilities (see these proceedings).

\section{Uncertainties in stellar structure and evolution}

 From observation the global stellar characteristics are the quantities more ``easily'' available (in particular luminosity and effective temperature), and eventually, the surface metallicity and gravity.
These can be reproduced using stellar models by adjusting global parameters like the age, the initial Helium abundance, and if not known - the mass.
But the key aspect is that the models have an extensive set of
options (mixing length parameter, relative abundances of specific heavy elements, extent of overshoot, additional mixing, the atmosphere, diffusion, to name a few) to build a particular solution that fully complies with the observed parameters.
These are all equally valid solutions unless further observational constraints can be found to discriminate and eliminate some.

\begin{figure}
\centering
\includegraphics[width=9cm]{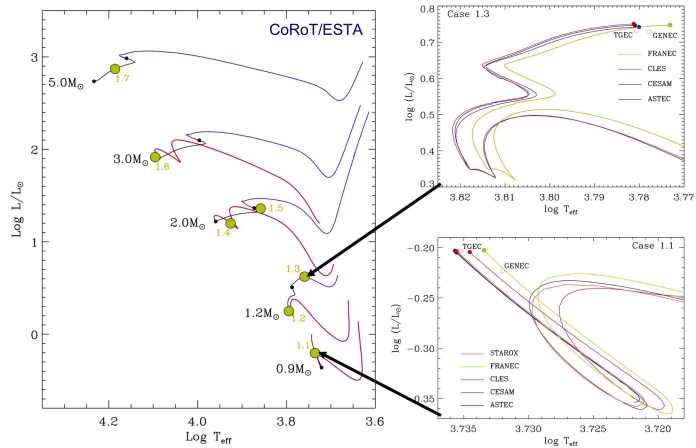}
\caption{Uncertainties/differences in the modelling of stellar structure and evolution evaluated under the CoRoT/ESTA effort using several evolution codes \citep{mjm-monteiro06}.
The interpretation of the observations is strongly dependent on the modelling leading to different conclusions for specific regions of the HR diagram where particular aspects of the physics may have an important role.}
\label{mjm-fig:models}
\end{figure}

The freedom to adjust several possible solutions is contained in (1) the prescriptions we select, (2) the uncertainties of our best prescriptions for calculating the internal structure of a star (see Fig.~\ref{mjm-fig:models}) and (3) the observational uncertainties of the stellar measurements.
There is such a wide range of competing uncertainties that it is very
difficult (some would say impossible) to be conclusive when reproducing the
photometric and astrometric observables.

In item (1) we also need to remind that there are dynamical effects on evolution usually not included, arising from the
complex interaction between convection (\citealt{mjm-straka06}), rotation (e.g. \citealt{mjm-roxburgh06}) and magnetic fields (including winds).
Contrary to standard current procedures, the effects of these components
of the physics on the evolution of the stars may have to be included
in the modelling, in particular for specific stars in the HR diagram.
But a detailed (fully validated) inclusion of those effects
in our models is still missing in most cases.

Regarding item (2) we may try to establish what are the uncertainties on the modelling and in particular in uncertainties introduced by numerical difficulties in adequately implement the physics on the models.
Some of these effects have been quantified but some aspects still need further improvement (see \citealt{mjm-monteiro06} for details).

High precision spectroscopy can help with item (3) as the range of allowed solutions can be significantly reduced if better, more precise stellar characteristics can be measured.
One example that requires improvement is the determination of the abundances in stars.
The models are extremely sensitive to these and no other observational data can replace the information content of a direct measurement of individual abundances of all relevant chemical species.

\section{Asteroseismic constraints}

Some known pulsators show the richness of oscillation data that we have found in our own Sun (e.g. \citealt{mjm-bedding07}).
With such a multitude of data it has become clear that much more
stringent constraints could be placed on the basic physical
ingredients used to model the structure and evolution of these stars.
If measured, oscillations frequencies increase greatly the available observational constraints on our models (e.g. \citealt{mjm-brown94,mjm-creevey07,mjm-eggenberger05}).

This makes Asteroseismology one of the most promising techniques for 
solving some of the long standing problems in Stellar Astrophysics.
The underlying goal is to be able to understand how some assumptions in
the modelling affect the observed properties, and so to use the
observed seismic behaviour to constrain those assumptions.
For a detailed discussion on how the frequencies can
complement the photometric, spectroscopic and astrometric observations, providing a
great improvement on tests of stellar models, see \citet{mjm-brown94}, \citet{mjm-cd99} or \citet{mjm-monteiro02} and references therein.

Different approaches within Asteroseismology have been developed
to infer specific constraints on the interior of different
type of pulsators.
Here we only mention some cases, in order to illustrate briefly the
type of information we can obtain from seismology which have a
direct consequence on the modelling of stellar structure and evolution.

A direct comparison of observed and model frequencies is usually the first step taken in using seismic data.
This is however a very difficult task as the potential for using an individual frequency depends strongly on knowing what type of mode it corresponds to.
This is in most cases very difficult, but high precision spectroscopy can be  a step forward by providing the data necessary for mode identification (see Aerts, these proceedings).
Some recent examples where it was attempted include the case of $\beta$~Cephei stars \citep{mjm-mazumdar06b} or young pre-main sequence stars \citep[and references therein]{mjm-ripepi06}.

The near surface structure of most stars is poorly know and our Sun has shown that it is inadequately represented in our models.
Consequently a direct comparison of frequencies is in most cases dominated by that aspect invalidating a test of the internal structure.
To overcome this difficulty several approaches have been developed based on the assumption that some adequate combinations of the frequencies can better test the interior structure without being very sensitive to what happens near the surface.
This is the approach being developed for studying pulsators with several identified frequencies of oscillation.
The Sun has shown us that the detail we may achieve is very high even if the data we expect for other stars is not as precise and abundant as what we have for our star.
The development of these tools is being actively extended (as an example see \citealt{mjm-monteiro02b,mjm-bazot05,mjm-mazumdar06a,mjm-cd07}) for studying other pulsators.
There is a wide variety of applications showing that frequencies can add an extremely important observational constraint allowing us to probe in detail the characteriostics of the stars being observed.
The detailed analysis by \cite{mjm-creevey07} illustrates the sensitivity of the parameters (quantities defining the models) to the observables (measured quantities).

An interesting example is the study of sub-giant stars which display solar-like oscillations.
As the star approaches the bottom of the red-giant region it becomes very difficult to differentiate stellar age and stellar mass using only luminosity and effective temperature.
If we add to the global parameters the oscillation frequencies the determination of stellar mass and age can be greatly improved allowing a more precise study of how the Helium core changes in this phase of evolution (e.g. \citealt{mjm-bedding06}).

Another set of very promising applications focus on the pre-main sequence evolution \cite{mjm-palla05,mjm-marques06,mjm-alecian07}.
Much can be gained from using seismology to study this complex phase of evolution in low and intermediate mass stars.

\section{Final remarks}

With the new observational opportunities provided by the space
missions in Asteroseismology, which will complement the coverage of
the HR Diagram with ground based observations, we may expect to
be able to address some of the unsolved problems in the modelling
of stellar structure and evolution.

The space missions in Asteroseismology: MOST launched in 2000 (\citealt{mjm-wlaker03}), CoRoT launched in 2006 (\citealt{mjm-baglin03}), and Kepler to be launched in 2008, will provide seismic data for several stars of different masses and in different phases of evolution.
All phases, from the pre-main sequence up to the pos-main sequence
evolution, can be cover complementing the already available data
from the ground on quite a few families of pulsating stars.
Additional efforts are being also prepared for ground based observations of which SONG is an example (\citealt{mjm-gundahl06} and these proceedings).

The much need increase in precision provided by a new generation of spectroscopic instruments (see these proceedings) on measuring stellar characteristics and oscillations are also expected to have a strong impact on Asteroseismology over the next decade.


\end{document}